# Election Bias: Comparing Polls and Twitter in the 2016 U.S. Election

**David Anuta, Josh Churchin & Jiebo Luo**

University of Rochester, Department of Computer Science
danuta@u.rochester.edu, jchurchi@u.rochester.edu, jluo@cs.rochester.edu

**Abstract**

While the polls have been the most trusted source for election predictions for decades, in the recent presidential election they were called inaccurate and biased. How inaccurate were the polls in this election and can social media beat the polls as an accurate election predictor? Polls from several news outlet and sentiment analysis on Twitter data were used, in conjunction with the results of the election, to answer this question and outline further research on the best method for predicting the outcome of future election.

## Introduction

Recently, polls have been called biased outlets for predictions due to the favoritism that their results seemed to exhibit towards Hillary Clinton, the Democratic candidate in the 2016 U.S. election. Twitter, on the other hand, has become a better-known source for election data. Consequently, two questions arise from the outcome of the U.S. presidential 2016 election: First, were the polls for the election (whether intentionally or not) biased, in the sense that they continuously and erroneously predicted a win for Hillary Clinton, and if they were, how much bias did they contain? Second, can social media be used to a greater effect as a more accurate, less biased, prediction model for the election or is it subject to the same bias/error that might be found within the polls? Our study aims to investigate both of these questions. We start by defining the term bias as we will use it within this paper.

When referring to election polls, a poll that is "biased" is one that incorrectly predicts the percentage of people that will vote for a given candidate. Polls can be either biased towards a candidate, meaning that they predict a candidate has a greater share of the vote than they actually do, or biased against a candidate, meaning that they predict a candidate has a smaller share of the vote than they actually do. The issue arises that colloquially, bias bring forth the concept of the *intentional* manipulation of data and/or an ideal towards one's own goals/agendas. When applied to polls, the colloquial definition of bias would bring with it a negative connotation that implies the conductor of the poll *intentionally* skewed the results towards an outcome that would reflect their political agenda. However, bias within a poll (as we have outlined above) does not necessarily speak to the *intentional* manipulation of data by the source.

Instead, it represents the difference between the data's prediction and reality. Thus, when viewing the numerical biases we represent throughout this paper, it should be noted that any bias represented is simply a quantification of accuracy of the data. The source of this inaccuracy, and whether it may have been intentional or not, is an entirely different topic which would require a much more in depth view into how each poll is conducted.

## Related Work

There have been several studies conducted based solely upon either bias in election polls or mining Twitter data to predict the outcome of an election. One study has drawn in depth conclusions about polling variance and biases in elections (Shirani-Mehr, Rothschild, Goel, and Gelman 2016). This literature is helpful in understanding the polling biases that may occur and possible numerical quantifications for this bias.

Studies have also debated and outlined several different models for predicting election outcomes using Twitter (Shi, Agarwal, Garg, Spoelstra 2012; Tumasjan, Sprenger, Sandner, Welpe 2010). We use the simplest method outlined in the related literature, using the number of users tweeting election related material about candidates as a guide. We combine this with outlined sentiment analysis techniques also discussed (Shi, Agarwal, Agarwal, Garg, Spoelstra 2012) to accurately predict the outcome of the election and get an accurate, unbiased result.

## Methodology

To answer the two central questions, we must examine the polls and Twitter from several different angles. The U.S.'s election process is thought to be a democratic one, but, in reality, its election process is actually republican due to the inclusion of the electoral college. Thus, in the U.S. election, there are two important battles which the polls and Twitter can predict: the overall popular vote and the electoral college (i.e. state) vote. We analyzed both these components using both poll and Twitter data and compared the each of the resulting biases.

For the popular vote, we collected various polls of the popular vote for the 2016 U.S. election from several different news sources. We then used Twitter users' tweets about the two major party candidates (Hillary Clinton and Donald Trump) to create a simple prediction model for the outcome of the popular vote. The two biases found using both these prediction methods were then compared to determine which, if either, was more biased when predicting the popular vote outcome.

For the electoral college, we restricted our in-depth analysis to nine different states for both the polls and Twitter. We chose four states that could be easily predicted from their very liberal or conservative learnings in the past (California, New York, Texas, and Tennessee). We also chose five states (Ohio, Wisconsin, Pennsylvania, Minnesota, Michigan) which were either: (a) viewed as very contentious "battleground" states, instrumental to Donald Trump's victory or (b) surprising swing states in this election. We analyzed poll data from only two sources due to the reduced amount of data available on a state by state basis. Resulting bias was compared with the bias we found using Twitter data mined for user location.

## Polls

**The Popular Vote** We gathered national popular vote percentage polls from eight different sources to use as our poll data. The timeframe for the polls ranged from as early as January 2016 until the day of the election, November 7th, 2016. We gathered at least 5 polls distributed over the course of the election from each of our 8 sources, allowing us to observe the source's bias over time. Our sources were: Fox News, Bloomberg/Selzer, Lucid/The Times-Picayune, YouGov/Economist, IBD (Investor's Business Daily), the New York Times/CBS News, NBC News/SurveyMonkey, and ABC News/The Washington Post. Many polls were conducted by several agencies working in conjunction with one another. For clarity, and to conserve space, we refer to these polls using only one of the two agency's names. Table 1 shows the last poll from each source conducted *right before* the results of the election.

| Source | Date of Poll | Trump % | Clinton % |
|---|---|---|---|
| YouGov | 11/04 - 11/07 | 41.0% | 45.0% |
| IBD | 11/04 - 11/07 | 43.0% | 41.0% |
| Bloomberg | 11/04 - 11/06 | 41.0% | 44.0% |
| Lucid | 11/04 - 11/06 | 40.0% | 45.0% |
| Fox News | 11/03 - 11/06 | 44.0% | 48.0% |
| ABC News | 11/03 - 11/06 | 43.0% | 47.0% |
| New York Times | 11/02 - 11/06 | 41.0% | 45.0% |
| NBC News | 10/31 - 11/06 | 41.0% | 47.0% |

*Table 1. Sample poll results from 8 outlet sources.*

This study focuses only on the two major parties in the U.S.. So, as is standard in the literature (Shirani-Mehr, Rothschild, Goel, and Gelman 2016), we adjusted the predicted percentages shown for each candidate to represent a two-party race between Hillary Clinton and Donald Trump. This was done using Equation (1) below:

$$AP_{C1} = \frac{P_{C1}}{P_{C1}+P_{C2}} \quad (1)$$

Where $AP_{C1}$ represents the adjusted percentage of the given candidate C1, $P_{C1}$ represents the non-adjusted percentage of candidate C1, and $P_{C2}$ represents the non-adjusted percentage of the second candidate C2. Sample results of this formula applied to the polls given in Table 1 are shown in Table 2.

| Source | Trump Adj. % | Clinton Adj. % |
|---|---|---|
| YouGov | 47.7% | 52.3% |
| IBD | 51.2% | 48.8% |
| Bloomberg | 48.2% | 51.8% |
| Lucid | 47.1% | 52.9% |
| Fox News | 47.8% | 52.2% |
| ABC News | 47.8% | 52.2% |
| New York Times | 47.7% | 52.3% |
| NBC News | 46.6% | 53.4% |

*Table 2. Sample data for adjusted candidate percentages from the poll data from Table 1. Adjusted percentage (Adj. %) for each candidate is calculated using formula (1).*

To obtain our measure of bias, we obtained the actual outcome of the popular vote for the 2016 election. As above, to represent a two-party race, we adjusted these percentages using formula (1). The resulting adjusted percentages (and original popular vote percentages for each candidate) are shown in Table 3.

| Candidate | Vote % | Adj. Vote % |
|---|---|---|
| Trump | 46.3% | 48.99% |
| Clinton | 48.2% | 51.01% |

*Table 3. Outcome of the popular vote from the 2016 U.S. election. Adj. Vote % is calculated for each candidate by using formula (1) on the vote percentage for both candidates.*

The sample data shown in Tables 1 and 2 is only a fraction of the data we collected. From each of our sources, we gathered at the very least 5 different polls, each at a different point in the election. An example of this time series data is shown in Table 4.

To produce a numerical measure of the overall bias for each of these sources, we aggregated our time series data for each of our sources into a single poll corresponding to a given month leading up to the election. For each source, we combined all the polls conducted in a given month into a single poll for that month to obtain a measure of the overall popular vote prediction in that given month by the source.

| Date of Poll | Num. Resp. | Trump % | Clinton % |
|---|---|---|---|
| 11/3 - 11/6 | 1295 | 44.0% | 48.0% |
| 11/1 - 11/3 | 1107 | 43.0% | 45.0% |
| 10/22 - 10/25 | 1221 | 41.0% | 44.0% |
| 10/15 - 10/17 | 912 | 39.0% | 45.0% |
| 10/10 - 10/12 | 917 | 38.0% | 45.0% |
| 10/3 - 10/6 | 896 | 42.0% | 44.0% |
| 9/27 - 9/29 | 911 | 40.0% | 43.0% |
| 9/11 - 9/14 | 867 | 40.0% | 41.0% |
| 8/28 - 8/30 | 1011 | 39.0% | 41.0% |
| 7/31 - 8/2 | 1022 | 39.0% | 49.0% |
| 6/26 - 6/28 | 1017 | 38.0% | 44.0% |
| 6/5 - 6/8 | 1004 | 39.0% | 42.0% |
| 5/14 - 5/17 | 1021 | 45.0% | 42.0% |
| 4/11 - 4/13 | 1021 | 41.0% | 48.0% |
| 3/20 - 3/22 | 1016 | 38.0% | 49.0% |
| 2/15 - 2/17 | 1031 | 42.0% | 47.0% |
| 1/4 - 1/7 | 1006 | 47.0% | 44.0% |

*Table 4. Polls conducted by Fox News over the course of the election. Each row is a poll from Fox News and contains the dates on which the poll took place (Date of Poll), the number of participants in the poll (Num. Resp.), and the candidate percentages for the given poll (Trump % and Clinton %).*

| Month | Trump % | Clinton % | Adj Trump % | Adj Clinton % |
|---|---|---|---|---|
| November | 43.5% | 46.6% | 48.3% | 51.7% |
| October | 40.1% | 44.5% | 47.4% | 52.6% |
| September | 40.0% | 42.0% | 48.8% | 51.2% |
| August | 39.0% | 45.0% | 46.4% | 53.6% |
| July | | | | |
| June | 38.5% | 43.0% | 47.2% | 52.8% |
| May | 45.0% | 42.0% | 51.7% | 48.3% |
| April | 41.0% | 48.0% | 46.1% | 53.9% |
| March | 38.0% | 49.0% | 43.7% | 56.3% |
| February | 42.0% | 47.0% | 47.2% | 52.8% |
| January | 47.0% | 44.0% | 51.6% | 48.4% |

*Table 5. Aggregated data for all polls collected by Fox News in each month (no polls were conducted in July). Raw percentages were calculated using formula (2). Adjusted percentages were calculated formula (1) and the calculated raw percentages.*

To create the non-adjusted monthly popular vote predictions for each source (as shown in Table 5 using the Fox News source) we applied Equation (2) to the polls of each source.

$$P_{m,C1} = \frac{\sum_{i=1}^{n} P_{i,C1} N_i}{\sum_{i=1}^{n} N_i} \quad (2)$$

Where $P_{m,C1}$ is the aggregated poll percentage for a candidate C1 in a given month $m$, which contains $n$ polls. Then, given a poll $i$ conducted in month $m$, $N_i$ is the number of respondents in poll $i$ and $P_{i,C1}$ is the percent that candidate C1 received in poll $i$. Thus, Equation (2) sums up the total number of people that decided to vote for candidate C1 in the given month, and divides it by the total number of people polled in that month to get the support percentage for candidate C1 over the course of the entire month. Once the support percentages are found, the percentage are adjusted using Equation (1) to represent a two-party race. Data obtained from using this process on the polls conducted by Fox News is shown in Table 5.

| Month | YouGov Bias | IBD Bias | Bloomberg Bias | Lucid Bias |
|---|---|---|---|---|
| November | -1.3% | 1.6% | -0.7% | -2.0% |
| October | -1.5% | 0.3% | -3.9% | -2.8% |
| September | -0.4% | | 1.0% | -2.4% |
| August | -2.3% | -1.3% | -1.4% | |
| July | -1.3% | | | |
| June | -1.4% | -1.7% | -6.0% | |
| May | 0.1% | | | |
| April | -0.8% | -3.0% | | |
| March | -0.4% | -6.3% | -9.0% | |
| February | | | | |
| January | | | | |
| Avg. Bias | -1.0% | -1.7% | -3.3% | -2.4% |

*Table 6. Monthly and average biases calculated for Trump from the first 4 of our 8 sources. A positive bias of 1.0% means that the source was biased towards Trump by that amount (i.e. it overpredicted the actual results of the election by that percentage). A negative bias of -1.0% means that the source was biased against Trump by that amount (i.e. it underpredicted the actual results of the election by that percentage).*

| Month | Fox News Bias | ABC News Bias | New York Times Bias | NBC News Bias |
|---|---|---|---|---|
| November | -0.7% | -1.0% | -1.2% | -2.7% |
| October | -1.6% | -0.8% | -0.6% | -2.7% |
| September | -0.2% | 1.0% | -0.1% | -2.1% |
| August | -2.6% | -2.7% | | -3.3% |
| July | | -1.2% | 1.0% | -0.9% |
| June | -1.8% | -5.7% | | -2.2% |
| May | 2.7% | 2.1% | -2.4% | -0.9% |
| April | -2.9% | | | -3.6% |
| March | -5.3% | -3.9% | -4.6% | -6.7% |
| February | -1.8% | | | |
| January | 2.7% | | | |
| Avg. Bias | -1.5% | -1.5% | -1.3% | -2.8% |

*Table 7. Monthly and average biases calculated for Trump from the first 4 of our 8 sources. See Table 6 for bias explanation.*

Equation (3) was used to find the monthly adjusted percentages of each source's polls.

$$B_{C1} = AP_{C1,s} - AP_{C1,f} \quad (3)$$

Where $B_{C1}$ is the bias towards candidate $C1$ of the poll being analyzed. The two terms, $AP_{C1,f}$ and $AP_{C1,s}$, are, respectively, the adjusted percentage of candidate $C1$ in the actual election and the adjusted percentage of candidate $C1$ in the poll currently being analyzed. Due to the nature of this formula and the adjusted percentages it is always the case that $B_{C1} = -B_{C2}$. Thus, for each poll we chose to only compute and show the bias towards or against Trump. The bias computed for each of the sources are shown in Tables 6 & 7. The average bias of each source over the course of the entire election (based on the data we collected) is shown on the last row of the tables.

**The Electoral College** We collected polls of the popular vote in each of our 9 states from two sources: The Washington Post/SurveryMonkey and UPI/CVoter. Formula (1) was then used on this data to allow for the analysis of a two-party race. Data collected for the 9 states and adjusted percentages are shown in Table 8.

| State | Source | Trump % | Clinton % | Adj Trump % | Adj Clinton % |
|---|---|---|---|---|---|
| *California* | UPI | 37.4% | 58.0% | 39.2% | 60.8% |
| | The Post | 29.0% | 48.0% | 36.4% | 63.6% |
| *New York* | UPI | 37.8% | 58.6% | 39.2% | 60.8% |
| | The Post | 31.0% | 51.0% | 37.8% | 62.2% |
| *Texas* | UPI | 54.1% | 40.4% | 57.2% | 42.8% |
| | The Post | 40.0% | 40.0% | 50.0% | 50.0% |
| *Tennessee* | UPI | 39.1% | 54.8% | 58.4% | 41.6% |
| | The Post | 31.0% | 51.0% | 62.2% | 37.8% |
| *Ohio* | UPI | 48.3% | 47.1% | 50.6% | 49.4% |
| | The Post | 40.0% | 37.0% | 51.9% | 48.1% |
| *Wisconsin* | UPI | 43.1% | 51.3% | 45.7% | 54.3% |
| | The Post | 37.0% | 39.0% | 48.7% | 51.3% |
| *Pennsylvania* | UPI | 47.1% | 47.8% | 49.6% | 50.4% |
| | The Post | 38.0% | 41.0% | 48.1% | 51.9% |
| *Minnesota* | UPI | 46.2% | 49.5% | 48.3% | 51.7% |
| | The Post | 34.0% | 41.0% | 45.3% | 54.7% |
| *Michigan* | UPI | 45.0% | 48.9% | 47.9% | 52.1% |
| | The Post | 38.0% | 39.0% | 49.4% | 50.6% |

*Table 8. State poll data from our two sources. Polls conducted by the Washington Post ran August 9th – September 1st and had anywhere from 500-5000 respondents. Polls collected from UPI ran November 4th – November 6th and contained anywhere from 300-500 respondents.*

We also collected the relevant states' actual outcome percentages for the 2016 election and adjusted their results using Equation (1). This data is show in Table 9. We used this data in conjunction with the poll data shown in Table 8 in Equation (3) to produce bias for each of the polls (shown in Table 10).

| State | Clinton % | Trump % | Adj. Clinton % | Adj. Trump % |
|---|---|---|---|---|
| *California* | 61.6% | 32.8% | 65.3% | 34.7% |
| *New York* | 58.8% | 37.5% | 61.1% | 38.9% |
| *Texas* | 58.8% | 37.5% | 61.1% | 38.9% |
| *Tennessee* | 34.7% | 60.7% | 36.4% | 63.6% |
| *Ohio* | 43.5% | 52.1% | 45.5% | 54.5% |
| *Wisconsin* | 46.9% | 47.9% | 49.5% | 50.5% |
| *Pennsylvania* | 47.6% | 48.8% | 49.4% | 50.6% |
| *Minnesota* | 46.9% | 45.4% | 50.8% | 49.2% |
| *Michigan* | 47.3% | 47.6% | 49.8% | 50.2% |

*Table 9. Actual results of 2016 election in each of the 9 states.*

| | UPI | Washington Post |
|---|---|---|
| *California* | 4.5% | 1.6% |
| *New York* | 0.3% | -1.1% |
| *Texas* | 2.5% | -4.8% |
| *Tennessee* | -5.3% | -1.4% |
| *Ohio* | -3.9% | -2.5% |
| *Wisconsin* | -4.9% | -1.8% |
| *Pennsylvania* | -1.0% | -2.5% |
| *Minnesota* | -0.9% | -3.9% |
| *Michigan* | -2.2% | -0.8% |

*Table 10. State poll biases. The left-hand column represents the state in which the poll was conducted. The top row represents the source that conducted the poll.*

### Twitter

**Data Collection** We used a dataset already retrieved from Twitter containing hundreds of millions of tweet ids (Littman et al. 2016). The collection contains tweet ids from tweets mined from July 13th until November 10th, filtered for material related specifically to either the election or to one of the two major candidates. Using Python and PostgreSQL, we retrieved as many of these tweets as was possible using the Twitter API in the time frame that we had.

To obtain an accurate prediction model for the US election, tweets were mined for unique users. These users in turn, were sorted by location within the US using some combination of tweet geotagging, user specified location, and a user's language preferences. All told, after location sorting, around 3 million tweets linked to roughly 750,000 unique user's remained. This data was stored in a PostgreSQL database to be used for sentiment analysis.

**Sentiment Analysis** Analysis was again conducted using Python and PostgreSQL. The analysis was conducted using the lexicon and rule-based sentiment analysis python library VADER sentiment (Hutto, Gilbert 2014). For each user, the text of each tweet associated with the user was analyzed using VADER to obtain a compounded sentiment probability (positive, negative, or neutral). Once analyzed the compound sentiment of a tweet contributed to the user's overall sentiment towards a candidate. A user has two compounded sentiments, one towards Hillary Clinton and one towards Donald Trump. The following criteria was used to determine how a tweet's sentiment should contribute to either or both of the user's overall compounded sentiments:

Tweet mentions single candidate[1] - *add compounded sentiment probability for that tweet to the user's overall sentiment towards that candidate.*

Tweet mentions both candidates – *add compounded sentiment probability for the tweet to both of the user's overall sentiments.*

Tweet mentions neither candidate – *do not update either of the user's compounded sentiments.*

A user's vote was defined as whichever candidate received a higher sentiment probability for that user.

To predict the popular vote using this data, each user's vote was counted. Of the nearly 750,000 users, a little over 250,000 of them had no vote associated with them, meaning that the users compounded sentiments towards both candidates were equal. The rest of the users successfully had a vote extrapolated from their compounded candidate sentiments. The number of users voting for each candidate was used to determine popular vote percentages by finding the number of users voting for a candidate and diving by the total number of users voting. Formula (2) was then used to the bias. The data found from the whole process is shown in Table 11.

|  | User Votes | User Vote % | Bias |
|---|---|---|---|
| *Trump* | 252011 | 52.4% | 3.4% |
| *Clinton* | 229346 | 47.6% | -3.4% |

*Table 11. Popular vote data based on Twitter. Extrapolated using sentiment analysis. User percent represents the number of users that were classified as for a candidate out of the total number of users classified. Bias is found using formula (2) and the actual popular vote outcome found in Table 3.*

The same process was used to predict state based popular voting, but for each state only users located in the state (as extrapolated from various user and tweet data) were used to find user voting percentages for each candidate. Table 12 shows the breakdown of the 9 states we chose.

## Analysis and Comparison

Comparisons of bias within this study are all based upon our computed numerical percentages of bias. These biases are percentages because they represent the overprediction (or underprediction) of the result being compared. For our analysis of both the popular vote and electoral college, we use the convention that a positive percentage bias means that a prediction overestimated Donald Trump's take of the popular vote by that many percentage points[2]. Thus, a positive bias represents a bias for Donald Trump, while a negative bias represents a bias against Donald Trump[3]. Unless otherwise stated, this should be the assumed convention for both our results and our comparisons.

| State | Candidate | User Votes | User Votes % | Bias |
|---|---|---|---|---|
| *California* | *Trump* | 11515 | 48.8% | 14.0% |
|  | *Clinton* | 12095 | 51.2% | -14.0% |
| *New York* | *Trump* | 9499 | 46.8% | 7.8% |
|  | *Clinton* | 10807 | 53.2% | -7.8% |
| *Texas* | *Trump* | 9676 | 53.6% | -1.2% |
|  | *Clinton* | 8373 | 46.4% | 1.2% |
| *Tennessee* | *Trump* | 2738 | 58.3% | -5.3% |
|  | *Clinton* | 1956 | 41.7% | 5.3% |
| *Ohio* | *Trump* | 4562 | 54.9% | 0.4% |
|  | *Clinton* | 3747 | 45.1% | -0.4% |
| *Wisconsin* | *Trump* | 1396 | 52.4% | 1.9% |
|  | *Clinton* | 1267 | 47.6% | -1.9% |
| *Pennsylvania* | *Trump* | 3945 | 50.8% | 0.1% |
|  | *Clinton* | 3827 | 49.2% | -0.1% |
| *Minnesota* | *Trump* | 1353 | 49.3% | 0.1% |
|  | *Clinton* | 1390 | 50.7% | -0.1% |
| *Michigan* | *Trump* | 2757 | 53.4% | 3.2% |
|  | *Clinton* | 2407 | 46.6% | -3.2% |

*Table 12. State based Twitter data. User Votes and User Votes % calculated as in Table 11 using users sorted by states. Bias calculated using formula (2) and actual results found in Table 9. Biases listed in the same row as Clinton are biases towards Clinton, meaning a positive bias represents a bias for her and a negative bias represents a bias against her.*

### Popular Voting Bias

**The Polls** Time series data, like what is shown in Table 4, was collected and graphed for each of our 8 sources[4]. The graphs, as well as a brief discussion of their similarities and differences are shown below. It should be noted that graphs are source specific, thus the poll dates (seen on the x-axis) are not standardized across all sources.

---

[1] Here mentioning a candidate meant that a tweet contained either the Democratic indicators 'Clinton' or 'Kaine' or the Republican indicators 'Trump' or 'Pence'.

[2] If the result of the election was that Donald Trump obtained 50% of the popular vote and the poll had a 1.0% bias, this would mean that the poll predicted that he would get 51% of the popular vote. Likewise, a bias of -1.0% means that the poll predicted he would get 49% of the popular vote.

[3] It should again be noted that the bias outline above is simply the deviation from the actual results. The fact that there is bias does not mean the bias was intentionally cultivated by the news source. A determination of a bias's source would require an in-depth consideration of the methods used in each poll, which is not done in this study.

[4] Only a single set of time series data is shown in table format in order to conserve space.

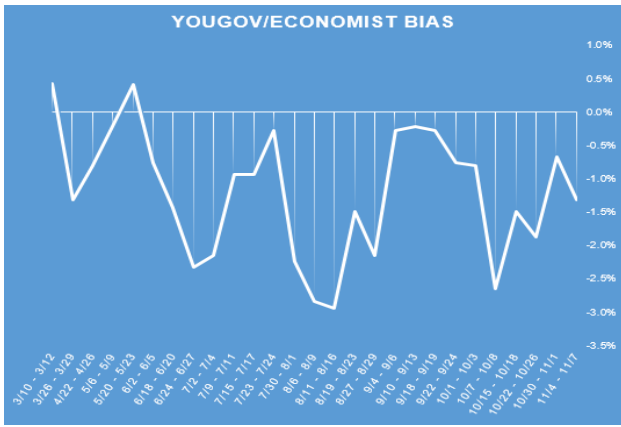

*Figure 1. A graph of bias for the polls conducted by YouGov/Economist poll. Dates for each poll conducted are on the x-axis and bias for each poll is on the y-axis.*

From Figure 1, it looks as if YouGov had a fairly steady bias against Trump throughout the election. Right before the election, a time when bias should be at its lowest point[5], YouGov has a small spike in bias against Donald Trump (from -0.7% to -1.3%). This trend is similarly reflected in several of our other sources.

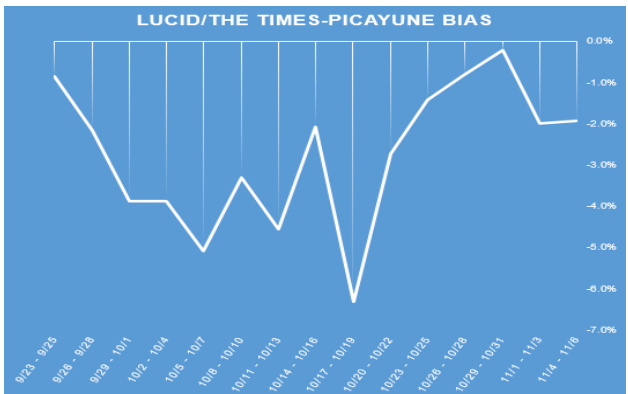

*Figure 2. A graph of bias for polls conducted by Lucid/The Times-Picayune. Dates for each poll conducted are on the x-axis and bias for each poll is on the y-axis.*

The range of bias for YouGov is the smallest of all the 8 sources we gathered, ranging only from 0.4% to −2.9% (range of 3.3% meaning that YouGov was always within 3.3% percentage points of predicting Trump's ending percentage). On average, YouGov had a -1.0% bias (Table 6) the lowest of all our poll sources.

---

[5] Bias should reduce over time as the election draws nearer. The country's feeling towards a candidate changes over time, so does the number of people that would vote for a certain candidate. The

Figure 2 clearly shows that Lucid was always biased against Trump from September up to election day. Their bias ranges from -0.2% to -6.3% (range of 6.1%) over time and overall their average bias of -2.4% (Table 6) was the third worst bias of our 8 sources. As with YouGov, Lucid also had an uptick in bias towards the end of the election.

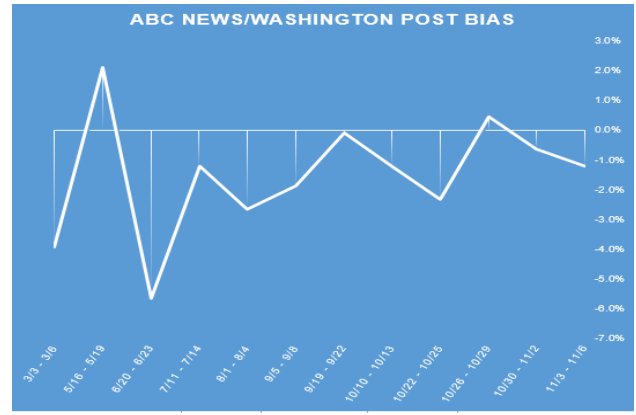

*Figure 3. A graph of bias for polls conducted by ABC News/The Washington Post. Dates for each poll conducted are on the x-axis and bias for each poll is on the y-axis.*

ABC is less consistently biased (Figure 3), but more varied over time with a range from −5.7% to 2.1% (range of 7.8%) over the entire election. After June, however, it was more consistent with a range from −2.7% to 0.5% (range of 3.2%), which is even more consistent than YouGov (range of 3.3%). Like with Lucid and YouGov, ABC also has an uptick in bias close to election day (from 0.5% to -1.2% over the course of a week). With an average bias of −1.5% (Table 7), ABC ties with Fox News for third place in terms of lowest biases of the sources that we collected.

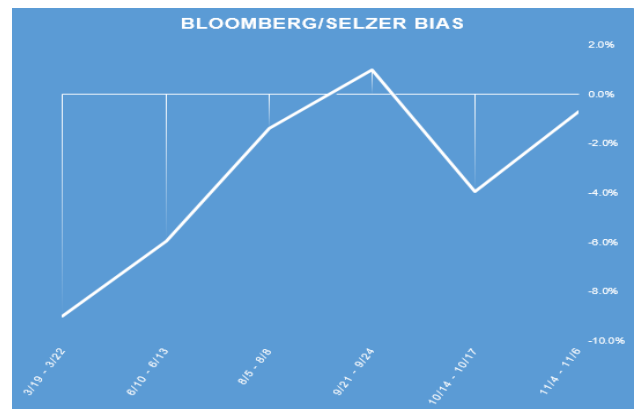

*Figure 4. A graph of bias for polls conducted by Bloomber/Selzer. Dates for each poll conducted are on the x-axis and bias for each poll is on the y-axis.*

closer a poll is conducted to the election, the more accurately it will reflect the feelings and opinions of the voters on election day.

The polls conducted by Bloomberg have biases ranging from -9.0% to 1.0% (range of 10%), the highest bias range of all 8 sources. As would be expected, this high range also corresponds to the worst average bias (-3.3% from Table 6) of all 8 sources. Despite the large starting bias (and thus range and average bias), the polls' biases clearly have a downward trend (Figure 4). Bloomberg tied with Fox News as the least biased source in November (right before the election) with a bias of -0.7%.

The New York Times (NYT) and Bloomberg have very similar bias curves (Figures 4 & 6), with NYT bias ranging from -4.6% to 1.0% (range of 5.6% as compared with Bloomberg's 10% range). Figure 6 shows a definite downward trend in bias right up until the election where a small uptick occurs (similar to ABC, Lucid, and YouGov). The small range and downward trend form an average bias of -1.3% (Table 7), the second lowest bias of the 8 sources, only coming in behind YouGov (average bias -1.0%).

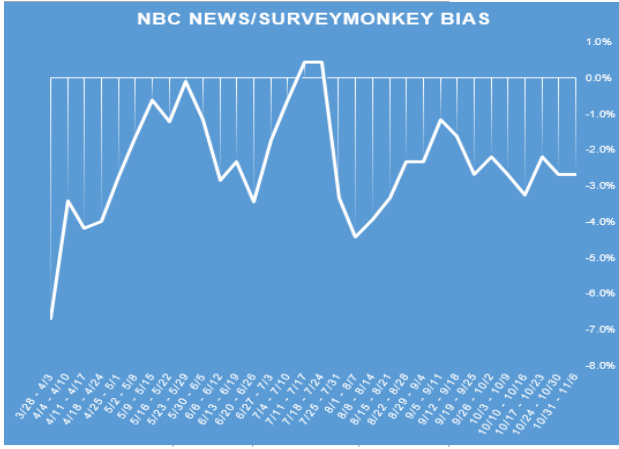

*Figure 5. A graph of bias for polls conducted by NBC News/SurveyMonkey. Dates for each poll conducted are on the a-axis and bias for each poll is on the y-axis.*

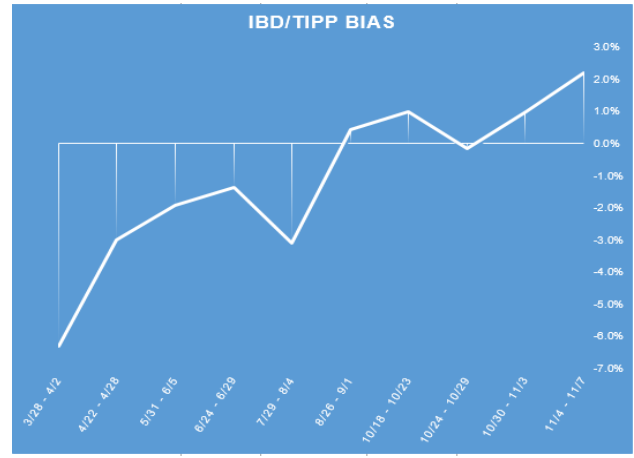

*Figure 7. A graph of bias for polls conducted by IDB/TIPP. Dates for each poll conducted are on the x-axis and bias is on the y-axis.*

NBC had the most polls conducted and the most participants in these polls. Despite this, Figure 5 shows their unstable bias ranging from -6.7% to 0.5% (range of 7.2%). This inconsistency and large range gives it an average bias of -2.8% (Table 7), the second worst bias of the 8 sources (the worst being Bloomberg). Figure 5 clearly shows that, unlike Bloomberg, there is no clear downward trend, but a consistent bias against Trump.

IDB was the only source (of the 8) which was biased towards Donald Trump by the end of the election. Its bias over time took on an almost linear shape (Figure 7), starting at -6.3% and ending at 2.2% (range of 8.5%). Like several other sources, in the last week preceding the election, an uptick in bias (−0.2% to 2.2%) occurred. Unlike all other sources, the uptick in bias occurred for Donald Trump. The decreasing nature of the bias until the week before the election gives IBD an average bias of −1.7% (Table 6).

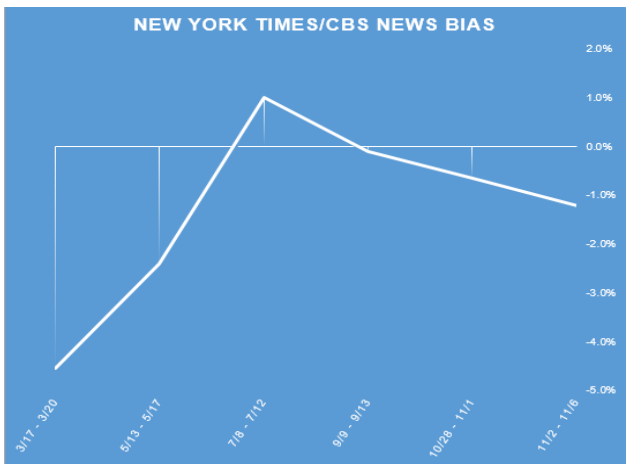

*Figure 6. A graph of bias for polls conducted by The New York Times/CBS News. Dates for each poll conducted are on the x-axis and bias is on the y-axis.*

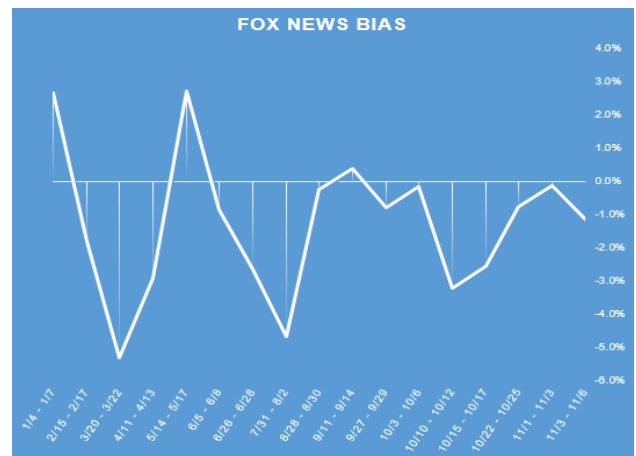

*Figure 8. A graph of bias for polls conducted by Fox News. Dates for each poll conducted are on the x-axis and bias for each poll is on the y-axis.*

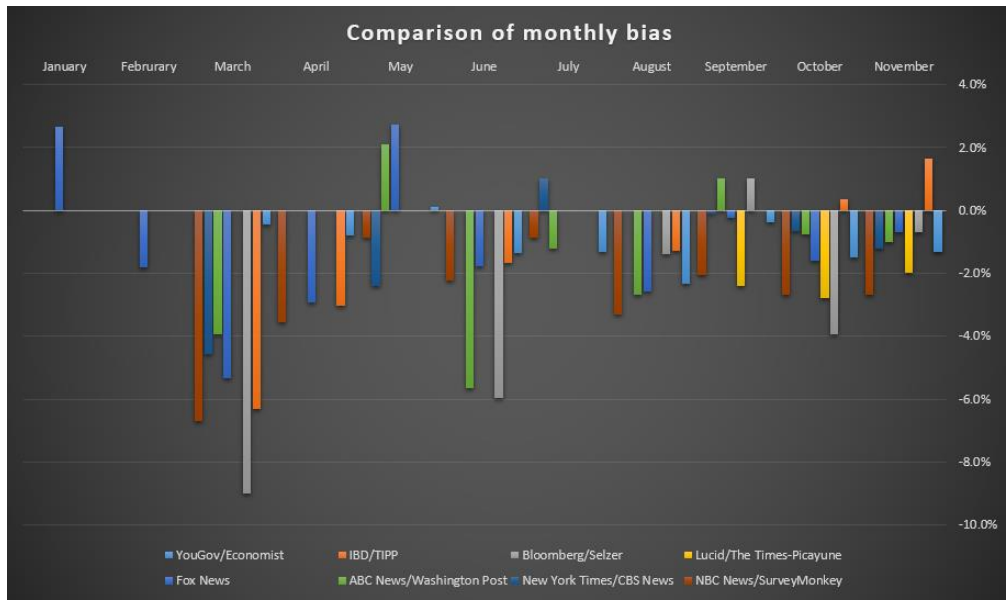

*Figure 9. A bar chart representing the monthly bias for each of the 8 source's biases. Data is taken from Tables 6 & 7 (minus the average bias contained in the bottom row).*

Fox News looks like the most volatile of the 8 sources (Figure 8) with biases ranging from -5.3% to 2.7% (range of 8.0%) in the short 2-month span from April to May. Interestingly, Fox News has a similar uptick in bias against Trump in the last week before the election (−0.1% to −1.2%). Average bias is -1.5% (Table 7) tied for third place with ABC News as third least biased. Unconventionally, average bias is against Donald Trump, despite the colloquial vision of Fox News as biased towards Republicans (and thus Donald Trump).

When comparing all 8 sources, there is noticeable reduction in bias range for all 8 sources over time (Figure 9). Most sources remained consistently biased against Donald Trump, with an average bias across all sources of −2.0%[6] over the whole election. In comparison, the average bias from all sources polls only 3 months before the election was -1.0%. Bias for Donald Trump was uncommon, even close to the election. Almost every source (save IBD) underpredicted the share of popular vote that Donald Trump would receive. Most of the sources, within three months of the election, had an average bias range of |1.3%|, with only Lucid and NBC having biases of over -2.0% (-2.4% and -2.5% respectively).

**Twitter vs the Polls** The sentiment analysis techniques applied to Twitter yielded a popular vote bias of 3.4% (Table 11) for Donald Trump. In comparison, this is 0.1% more biased (in the opposite direction) than our most biased poll source: Bloomberg, which had a bias against Donald Trump of -3.3% (Table 6). On average for all 8 sources, poll data had a bias of -2.0%. Comparatively, Twitter had a larger bias in the opposite direction with its 3.4% for Donald Trump. When compared with the poll biases in the 3 months leading up to the election, Twitter performed far worse than the polls which had an average bias of -1.0% over this 3-month period. Lucid's -2.8% bias in October was the closest to Twitter's 3.4% percent bias over this period.

**Electoral College Bias**

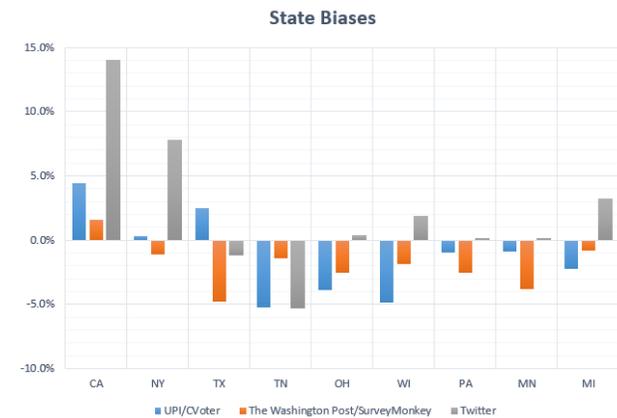

*Figure 10. Bar chart biases from both the polls and Twitter organized by state. The y-axis is the poll bias and the x-axis is the state.*

---

[6] Most sources predicted, on average across the whole election, that Donald Trump would get 2.0% less of the popular vote than he actually did.

**Poll Bias** Biases within states polls were somewhat volatile (Figure 10). Over the 9 states analyzed, the average biases for the Washington Post and UPI were -2.1% and -1.2% respectively. Biases within California, Texas, and Tennessee were all greater than 1.3% (in either direction), with New York being the only state in this group of easily predicted states that had relatively low biases (0.3% and −1.1% for UPI and The Post respectively). Biases across swing states Ohio, Wisconsin, Pennsylvania, Minnesota, and Michigan varied in their range (as low as -0.8% in Minnesota and as high as -4.9% in Wisconsin), but not in their direction. All swing state polls had biases against Donald Trump. Average poll biases across all states, for both sources, was -1.6%.

**Twitter vs. the Polls** Twitter's biases across the 9 states varied widely as seen in Figure 10. Compared with the polls' average of -1.6%, Twitter had a higher average bias of 2.4%. Twitter's highest biases of 14%, 7.8%, and -5.3% (for California, New York, and Tennessee respectively) were in states that should have been easier to predict. Twitter had some very low biases in Ohio, Pennsylvania, and Minnesota (under 0.5%), which were all considered harder to predict.

# Findings

## Bias in the polls

The first question we try to answer is whether there was any truth to the accusations that the 2016 U.S. election polls were biased towards Hillary Clinton. Our results show that these accusations have some grounding in the truth.

### The Popular Vote

Across our 8 sources, every single one had an average bias against Donald Trump throughout the election. On average, all our poll sources had a -2.0% bias, with YouGov having the lowest average bias (-1.0%) and Blommberg having the highest average bias (-3.3%). Over the course of the last 3 months leading up to the election this bias decreased (as it should) to an average bias of only −1.0%, with ABC having the lowest average bias (-0.3%) and NBC News having the highest average bias (-2.5%).

### The Electoral College

The electoral college or state based bias within the polls also consistently underpredicted Donald Trump's share of the popular vote. Overall, UPI and The Washington Post had an average bias of -1.6% across California, New York, Texas, Tennessee, Ohio, Wisconsin, Pennsylvania, Minnesota, and Michigan. Interestingly, the polls had large biases on three of the four states (California, Texas, and Tennessee) which should have been easily predicted and in California both poll sources were actually biased towards Donald Trump.

To answer our question, it seems clear from our data that all our sources, over the course of the 2016 US election, had a slight bias towards Hillary Clinton (and thus against Donald Trump) both in the popular vote and the electoral college. Quantifiably, all the polls, including Fox News (which is known colloquially for its bias towards Republican candidates) had average biases against Donald Trump. Overall, our 8 sources had a -2.0% average bias when predicting the popular vote and our 2 sources had a −1.6% average bias when predicting the electoral college.

## Social Media Bias

The second question we tried to answer in this study was whether Twitter could be a less biased predictor of the 2016 US election than the polls. Our results indicate that the technique we used actually results in a more biased predictor for both the popular vote and the electoral college.

### The Popular Vote

Twitter predicted that Hillary Clinton would win 47.6% of the popular vote and Donald Trump would win 52.4% of popular vote. The actual popular vote percentages for each candidate (adjusted for a two-party race) were 49.0% and 51.0% for Trump and Clinton respectively. We found that this equated to a 3.4% bias towards Donald Trump, meaning that Twitter predicted Donald Trump would win 52.4% of the popular vote versus the 49.0% that he did win. As compared with the poll's average bias of -2.0%, social media performed far worse in terms of absolute bias and was biased towards, instead of against, Donald Trump.

### The Electoral College

Twitter showed very little bias when it came to predicting the three swing states of Ohio (0.4%), Pennsylvania (0.1%), and Minnesota (0.1%). When predicting the easily predicted states of California, New York, and Tennessee twitter showed much higher biases of 14%, 7.8%, and −5.3% respectively. It's biases for Wisconsin (1.9%), Michigan (3.2%), and Texas (-1.2%) were similar to poll biases across all states. Across these 9 states, Twitter had an average bias of 2.4% towards Donald Trump.

To answer our question, Twitter had higher absolute bias than the polls in both the popular vote and the state based results, meaning that it was a more biased predictor for the election. The difference in bias is not only absolute but candidate sensitive. While the polls were biased against Donald Trump by -2.0% in the popular vote and -1.6% in the state based vote, Twitter was biased towards Donald Trump by 3.4% in the popular vote and 2.4% in the state based vote.

## Conclusion

In conclusion, we found that in the 2016 U.S. election the media (as encapsulated by our 8 sources) was, quantifiably, biased against Donald Trump by -2.0% in the popular vote and -1.6% in the state based votes over the entire election period. Towards the end of the election (in the 3-month period before election day) the popular vote bias decreased slightly to a -1.0% bias against Donald Trump. These numerical biases seem to point to the fact that our 8 sources were biased towards Hillary Clinton.

Interestingly enough, Twitter was more biased in the both the popular vote and the state based results with a 3.4% bias in the popular vote and a 2.4% bias in the state based results. Twitter's bias, unlike poll bias, was for Donald Trump instead of against him. Overall, Twitter performed worse than the polls when it came to predicting both the popular vote and the state based votes[7]. Thus, overall Twitter was more biased than the polls, but for a different candidate.

Both the polls and Twitter were biased in the 2016 U.S. election. Their biases not only differed in their quantity, but also in their direction. While the polls had a small bias towards Hillary Clinton, Twitter had a slightly larger bias towards Donald Trump. It is important to remember the difference, described at the beginning of this study, between the numerical "bias" we have described and the colloquial use of the term to indicate the intentional manipulation of data towards an agenda. The source of the bias described in this study is not addressed by this paper.

## Future Work

Our study has, in no way, been entirely exhaustive of this topic. Social media continues to be possible medium for election prediction, however, more studies should be done that expand upon this research to determine if there are any benefits to election prediction through social media as opposed to the standard method of using polls. There are several different ways this topic could be expanded.

One such future exploration should look to encompass more data. We only used eight sources of poll data and were only able to mine and use 800,000 Twitter users. Increasing either (or both) the number of poll sources (and/or polls from these sources) or the number of users classified would help to further this investigation of bias within the polls and social media.

Along with using more data, this topic could be expanded to utilize different methods of extracting election predictions from social media. Applying different sentiment analysis techniques or using different social media platforms may change the bias of social media. Comparing the biases of different techniques of social media election prediction could help to point towards the least biased technique.

---

[7] For the state based results, Twitter performed incredibly well in very close races. However, in states which heavily favored one candidate over the other, Twitter performed very poorly.